\documentclass[10pt,journal,cspaper,compsoc]{IEEEtran}
\ifCLASSOPTIONcompsoc
\else
   \usepackage{cite}
\fi
\usepackage{epsfig,multirow}

   \graphicspath{{.}}
  \DeclareGraphicsExtensions{.eps}   

\usepackage{amsfonts}
\ifCLASSOPTIONcompsoc
\usepackage[tight,normalsize,sf,SF]{subfigure}
\else
\usepackage[tight,footnotesize]{subfigure}
\fi

 \let\MYorigsubfigure\subfigure
 \renewcommand{\subfigure}[2][\relax]{\MYorigsubfigure[]{#2}}

\usepackage{url}

\begin{document}
%
\title{Semantic Computing of Moods Based on Tags in Social Media of Music}
%
%
%
%

\author{Pasi~Saari and
        Tuomas~Eerola
\IEEEcompsocitemizethanks{\IEEEcompsocthanksitem P. Saari and T. Eerola are with the Department
of Music, University of Jyv\"askyl\"a, Jyv\"askyl\"a,
Finland.\protect\\
E-mail: pasi.saari@jyu.fi.}
\thanks{This work was funded by the Academy of Finland (Finnish Centre of Excellence in Interdisciplinary Music Research).}}

%
%

\markboth{IEEE TRANSACTIONS ON KNOWLEDGE AND DATA ENGINEERING}%
{Saari \MakeLowercase{\textit{and}} Eerola: Semantic Computing of Moods Based on Tags in Social Media of Music}
%


\IEEEcompsoctitleabstractindextext{%
\begin{abstract}
Social tags inherent in online music services such as Last.fm provide a rich source of information on musical moods. The abundance of social tags makes this data highly beneficial for developing techniques to manage and retrieve mood information, and enables study of the relationships between music content and mood representations with data substantially larger than that available for conventional emotion research. However, no systematic assessment has been done on the accuracy of social tags and derived semantic models at capturing mood information in music. We propose a novel technique called Affective Circumplex Transformation (ACT) for representing the moods of music tracks in an interpretable and robust fashion based on semantic computing of social tags and research in emotion modeling. We validate the technique by predicting listener ratings of moods in music tracks, and compare the results to prediction with the Vector Space Model (VSM), Singular Value Decomposition (SVD), Nonnegative Matrix Factorization (NMF), and Probabilistic Latent Semantic Analysis (PLSA). The results show that ACT consistently outperforms the baseline techniques, and its performance is robust against a low number of track-level mood tags. The results give validity and analytical insights for harnessing millions of music tracks and associated mood data available through social tags in application development.


\end{abstract}

}

\maketitle

\IEEEdisplaynotcompsoctitleabstractindextext

%
\IEEEpeerreviewmaketitle

\section{Introduction}

\IEEEPARstart{M}{ining} moods inherent in online content, such as web forums and blogs \cite{Abbasi2008,Nguyen2013}, images \cite{Schmidt2009}, and news stories \cite{Bao2012}, brings benefits to document categorization and retrieval due to the availability of large data. The need for automatic mood-based music management is increasingly important as music listening, consumption and music-related social behaviors are shifting to online sources, and a large proportion of all recorded music is found online. An extensive body of research in music psychology has shown that moods\footnote{In this paper, we use terms mood, emotion, and affect interchangeably.} are, in many aspects, fundamental to music \cite{Juslin2009handbook}: music expresses and evokes moods, appeals to people through moods, and is conceptualized and organized according to moods. Online music services based on social tagging, such as Last.fm,\footnote{Last.fm: \url{http://www.last.fm/}.} exhibit rich information about moods related to music listening experience. Last.fm has attracted wide interest from music researchers, since crowd-sourced social tags enable study of the links between moods and music-listening in large music collections; these links have been unattainable in the past research, which has typically utilized laborious survey-based annotations.

Social tags can be defined as free-form labels or keywords collaboratively applied to documents by users in online services, such as Del.icio.us (web bookmarks), Flickr (photos), and Pinterest (images, videos, etc.)\footnote{Del.icio.us: \url{http://www.delicious.com}; Flickr: \url{http://www.flickr.com}; Pinterest: \url{http://pinterest.com}.}. Obtaining semantic information from social tags is, in general, a challenge not yet met. Due to the free-form nature of social tags, they contain a large amount of user error, subjectivity, polysemy and synonymy \cite{Golder2006}. In particular, the sparsity of social tags, referring to the fact that a typical document is associated to only a subset of all relevant tags, is a challenge to the indexing and retrieval of tagged documents. Various techniques in semantic computing, such as Latent Semantic Analysis (LSA) \cite{Deerwester1990}, that infer semantic relationships between tags from within-document tag co-occurences, provide solutions to tackle these problems, and techniques have been proposed to automatically predict or recommend new tags to documents bearing incomplete tag information \cite{Heymann2008,Song2011}. However, no agreement exists on how to map tags to the space of semantic concepts, as indicated by the large number of approaches dedicated to the task \cite{Garcia-Silva2012}.

In the music domain, the majority of social tags are descriptors of the type of music content, referring typically to genres \cite{Bischoff2008}, but also to moods, locales and instrumentations, which are well represented in the data as well. In particular, moods are estimated to account for 5\% of the most prevalent tags \cite{lamere2008}. Several studies in the field of Music Information Retrieval (MIR) have applied bottom-up semantic computing techniques, such as LSA to uncover mood representations emerging from the semantic relationships between social tags \cite{Levy2007,laurier2009b}. These representations have resembled mood term organizations in the dimensional \cite{Russell1980} or categorical \cite{Zentner2008,Ekman1992} emotion models, which have regularly been used to model moods in music \cite{Eerola2012}. However, we claim that the previous studies in tag-based music mood analysis have not given comprehensive evaluation of the models proposed, utilized knowledge emerging from emotion modeling to the full potential, or presented systematic evaluation of the accuracy of the models at the track level.

In this paper we propose a novel technique called Affective Circumplex Transformation (ACT), optimized to uncover the mood space of music by bottom-up semantic analysis of social tags. The key aspect of ACT is that it is a predictive model that can be used to predict the expressed moods in novel tracks based on associated tags. We train ACT with a large collection of approximately 250,000 tracks and associated mood tags from Last.fm and evaluate its predictive performance with a separate test set of 600 tracks according to the perceived moods rated by a group of participants. We compare ACT to predictive models devised based on various semantic analysis techniques, as well as to the predictions based on raw tag data. We also estimate the applicability of ACT to large collections of weakly-labeled tracks by assessing ACT performance as a factor of the number of tags associated to tracks. Furthermore, we gain insights into the general views on mood modeling in music by examining the structure of the mood semantic space inherent in social tags.

The rest of the paper is organized as follows: Section \ref{sec:related} goes through related work in semantic computing and emotion modeling. Section \ref{sec:method} describes the process of obtaining tracks and associated social tags from Last.fm and details the method for semantic analysis of the data. The semantic structures of the data are examined in Section \ref{sec:hopkins}. Section \ref{sec:act} presents the ACT technique and Section \ref{sec:baseline} introduces the baseline techniques for comparatively evaluating its prediction performance on listener ratings of the perceived mood in music. The test set used in the evaluation is described in Section \ref{sec:gt}. The results are presented and discussed in Section \ref{sec:results} and conclusions are drawn in Section \ref{sec:conclusions}.

\section{Related Work}\label{sec:related}

\subsection{Semantic Analysis of Social Tags}

Latent Semantic Analysis (LSA) \cite{Deerwester1990}, has been widely used to infer semantic information from tag data. To enable computational analysis, tag data is first transformed into the Vector Space Model (VSM) \cite{Salton1975}, representing associations between documents and tags in a sparse term-document matrix. Semantically meaningful information is then inferred from a low-rank approximation of the VSM, alleviating the problems with synonymy, polysemy and data sparsity. Low-rank approximation is typically computed by Singular Value Decomposition (SVD), but other techniques such as Nonnegative Matrix Factorization (NMF) \cite{Lee2001} and Probabilistic Latent Semantic Analysis (PLSA) \cite{Hofmann2001} have been proposed for the task as well. 

SVD has been used in past research for music auto-tagging \cite{Law2010} and music mood modeling \cite{Levy2007,laurier2009b}. Variants of NMF have been exploited for collaborative tagging of images \cite{Zhou2011} and user-centered collaborative tagging of web sites, research papers and movies \cite{Peng2011}. PLSA has been used for collaborative tagging of web sites \cite{Wetzker2009} and topic modeling of social tags in music \cite{Levy2008}.  In the latter paper, SVD and PLSA were compared in a task of genre and artist retrieval based on social tags for music, showing the advantage of PLSA in these tasks. Performance of SVD and NMF were compared in \cite{Rakesh2009}, in a bibliographic metadata retrieval task, but no significant difference was found. On the other hand, NMF outperformed SVD and PLSA in classification of text documents into mood categories \cite{Calvo2012}.

\subsection{Structure of Moods}\label{sec:moodspsych}

Emotion modeling in psychology and music psychology research typically relies on explicit -- textual or scale-based -- participant assessments of emotion term relationships \cite{Russell1980,Scherer1984,Thayer1989} and their applicability to music  \cite{Juslin2004,Zentner2008,juslin2011emotional}. Based on these assessments, dimensional \cite{Russell1980} and categorical \cite{Ekman1992} models of emotions have been proposed. Categorical emotion models either stress the existence of a limited set of universal and innate basic emotions \cite{Ekman1992}, or explain the variance between moods by means of a few underlying affect dimensions \cite{Thayer1989} or a larger number of emotion dimensions based on factor analyses \cite{Zentner2008}. With regards to music, an ongoing related theoretical debate considers whether moods in music can most realistically be described as categories or dimensions \cite{zentner2010a}. Two variants of the dimensional models of emotions \cite{Russell1980,Thayer1989} are particularly interesting here since these have received support in music-related research \cite{Eerola2012}. Russell's \cite{Russell1980} affective circumplex postulates two orthogonal dimensions, called Valence and Arousal, and these dimensions are thought to have distinct physiological substrates. Thayer's popular variant \cite{Thayer1989} of this dimensional model assumes the two dimensions to be rotated by $45^{\circ}$, labeling them as Tension and Energy. However, divergent views exist as to whether two dimensions is enough to represent affect. In particular, a three-dimensional model of Sublimity, Vitality and Unease has been proposed as underlying dimensions of affect in music \cite{Zentner2008}, whereas a model of Arousal, Valence and Dominance has been proposed as a normative reference for English words \cite{Bradley1999}.

Importantly, these models lend themselves to a coherent spatial representation of the individual affect terms, which is valuable property with respect to semantic analysis of mood-related social tags. 

Past accounts of mood detection in MIR have applied the various emotion models coupled with advanced techniques of machine learning and signal processing to identify acoustic substrates for moods. Both categorical \cite{Saari2011} and dimensional \cite{Yang2008} models of emotions have been used to represent the mood in music tracks. These studies prove that insights and findings from emotion modeling research are useful to new computational approaches to automatic mood modeling. Moreover, and as noted above, past studies have recovered mood spaces based on semantic analysis of social tags that resemble the emotion models \cite{Levy2007,laurier2009b}. Here, we go further by quantifying the predictive value of applying insights from the psychology of emotion to the analysis of large-scale and diffuse meta-data, such as information provided by social tags.

\begin{figure*}[tp!]
\centering
\includegraphics[width=1.5\columnwidth]{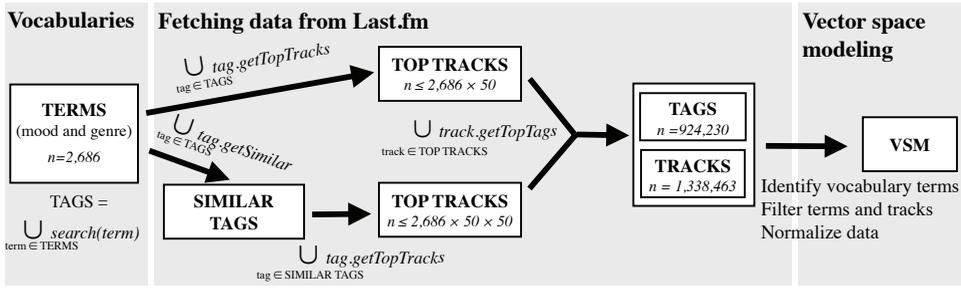}
\caption{Data collection process.}
\label{fig:graph}
\end{figure*}

\section{Semantic Analysis}\label{sec:method} 

This section describes the process of collecting tracks and associated social tags from Last.fm, and details the semantic analysis method to infer spatial representation of tags.

\subsection{Gathering Vocabularies}\label{subsec:reflist}

To obtain the corpus of tracks and associated tags from Last.fm, we systematically crawled the Last.fm online database through a dedicated API\footnote{\url{http://www.last.fm/api}, accessed during November - December 2011.}. We used extensive vocabularies of mood- and music genre-related terms as search words for populating the corpus. This approach suits our purposes since it controls the relevance of the track content and to some degree balances the data according to mood and genre.

The mood vocabulary was aggregated from several research papers and from an expert-generated word list. A number of research fields provided relevant sources: affective sciences \cite{shaver1987}, music psychology studying the use of emotion words in music \cite{Zentner2008,hevner1936,Russell1980} and MIR \cite{HuDownie2007,laurier2009b}, studying the mood prevalence in social tags. As an expert-generated source we used an extensive mood word list at Allmusic\footnote{\url{http://www.allmusic.com/moods}} web service. The vocabulary was then edited manually to identify inflected terms, such as ``depressed'', ``depressing'', ``depression'' and ``depressive''.

Genre vocabulary was aggregated from several expert-generated sources. A reference for music genres and styles available through Allmusic\footnote{\url{http://www.allmusic.com/genres}} was used as the main source. This included over 1,000 popular and classical music styles. Moreover, we included several Finnish music styles out of curiosity. Manual editing was then carried out for the genre vocabulary to aggregate regular alternate spellings, such as ``rhythm and blues'',  ``R'n'B'', and ``R\&B'' as well as ``indie pop'' and ``indiepop''.

The number of terms in the resulting vocabularies was 568 for moods and 864 for genres (1,083 and 1,603 including the inflected forms, respectively). 

Moreover, the following reference vocabularies were collected for evaluating the mood structures in Section \ref{sec:hopkins}:
\textbf{Locations -- 464 terms:} Country names including nationality-related nouns and adjectives (e.g., ``Finland'', ``Finn'', ``Finnish''), as well as continents and certain geographical terms (e.g., ``arctic'').
\textbf{Instruments -- 147 terms:} Comprehensive list of instrument names.
\textbf{Opinions -- 188 terms:} Manually identified from the tags associated to more than 1,000 tracks, and not included in the other vocabularies (e.g., ``favorite'', ``one star'', ``wicked'', ``check out'').

\subsection{Fetching Tags and Tracks from Last.fm}\label{subsec:tags}

The mood and genre vocabularies, including the inflected terms, were used as search words via the Last.fm API\footnote{Find detailed information on the used functions from the API documentation referenced above.} to populate the track corpus. The process is visualized in Fig. \ref{fig:graph}.

The tracks were collected using two tag-specific API functions: \textit{tag.getTopTracks} returning up to 50 top tracks and \textit{tag.getSimilar} returning up to 50 most similar tags. First, the top tracks for each term were included in the corpus, amounting to up to $2,686 \times 100 = 134,300$ tracks. In parallel, for each term we fetched the similar tags and included the associated top tracks. This process potentially visited up to $2,686 \times 50 \times 50 = 6,715,000$ tracks, and using both fetching processes combined we were able to fetch up to approximately 7M tracks. In practice, the number was reduced by many overlapping tracks and similar tags.

Finally, track-level tags in the final corpus were fetched using the function \textit{track.getTopTags}, returning up to 100 tags. The returned track-level tags are represented by normalized ``counts'' indicating the relative number of times each tag has been applied to a track. Although the exact definition of these counts is not publicly available, they are often used in semantic analysis \cite{lamere2008,laurier2009b}. All tags were cleaned by lemmatizing \cite{Fellbaum1998} and by removing non-alphanumeric characters. The final set consisted of 1,338,463 tracks and 924,230 unique tags.

\subsection{Vector Space Modeling}

A standard Vector Space Model (VSM) \cite{Salton1975} was built separately for each of the vocabularies. Tags related to the vocabulary terms were identified from the corpus following the bag-of-words approach also taken in \cite{Levy2008}. All tags that included a term as a separate word (or separate consecutive words in the case of multi-word terms) were associated with the corresponding terms. We also filtered out those track-specific term associations where a term was included in either track title or artist name. This was due to the fact that many social tags describe these actual track metadata.

\begin{figure}[!t]
\includegraphics[trim=11mm 0mm 16mm 0mm, clip, width=\columnwidth]{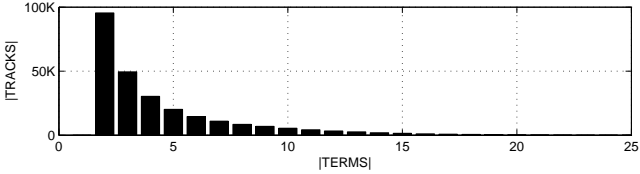}%
\caption{Distribution of the number of mood terms associated to tracks in the test set.}
\label{fig:tagsums}
\end{figure}

To avoid obtaining overly sparse and uncertain information, we excluded all terms that were associated to less than 100 tracks. At this point 493,539 tracks were associated to at least one mood term. However, we excluded tracks associated to only one mood term, as it was assumed that these tracks would provide little additional information for the further semantic analysis of mood term relationships. This resulted in a corpus of 259,593 tracks and 357 mood terms. As shown in Fig. \ref{fig:tagsums}, distribution of the number of terms associated to each track was exponential, indicating the sparsity of the data. Similar procedures were applied to all other vocabularies as well. Statistical measures related to the resulting corpora are shown in Table \ref{table:vectorspace}. The five most frequently applied within each corpora are as follows: \textbf{Moods}: ``chill'', ``mellow'', ``relaxing'', ``dark'' and ``melancholy''; \textbf{Genres}: ``rock'', ``pop'', ``alternative'', ``electronic'' and ``metal''; \textbf{Instruments}: ``guitar'', ``bass'', ``drum'', ``piano'' and ``acoustic guitar''; \textbf{Locales}: ``British'', ``UK'', ``American'', ``USA'' and ``German''; and \textbf{Opinions}: ``favorite'', ``love'', ``beautiful'', ``awesome'' and ``favourite''.

Finally, the normalised counts $n_{i,j}$ provided by Last.fm for term ($w_i$) -- track ($t_j$) associations were used to form the VSM $N$ defined by Term Frequency-Inverse Document Frequency (TF-IDF) weights $\hat{n}$ in a similar manner as in \cite{Levy2008}:
\begin{equation}\label{eq:tfidf}
\hat{n}_{i,j}=(n_{i,j} +1)\log(\frac{R}{f_i}),
\end{equation}
where $R$ is the total number of tracks, $f_i$ is the number of tracks term $w_i$ has been applied to. Separate models were formed for each vocabulary-related corpora.

\begin{table}[!t]
\caption{Statistical measures related to each vocabulary-related corpus.}
\label{table:vectorspace}
\centering
\begin{tabular}{l|l|l|l}
&$|$Tracks$|$&$|$Terms$|$&\# Terms per track (Avg.)\\
\hline
Moods&259,593&357&4.44\\
Genres&746,774&557&4.83\\
Instruments&46,181&53&4.44\\
Locales&72,229&126&2.26\\
Opinions&305,803&192&5.67\\
\hline
\end{tabular}
\end{table}

\subsection{Singular Value Decomposition}\label{subsec:semantics}

SVD is the typical low-rank matrix approximation technique utilized in LSA to reduce the rank of the TF-IDF matrix, alleviating problems related to term synonymy, polysemy and data sparsity. SVD decomposes a sparse matrix $N$ so that $N = USV^T$, where matrices $U$ and $V$ are orthonormal and $S$ is the diagonal matrix containing the singular values of $N$. Rank $k$ approximation of $N$ is computed by $N^k = U^kS^k(V^k)^T$, where the $i$:th row vector $U^k_i$ represents a term $w_i$ as a linear combination of $k$ dimensions. Similarly, $V^k_j$ represents track $t_j$ in $k$ dimensions. Based on a rank $k$ approximation, dissimilarity between terms $w_i$ and $w_{\hat i}$ is computed by the cosine distance between $U^k_iS^k$ and $U^k_{\hat i}S^k$.

In the present study, all data sets summarized in Table \ref{table:vectorspace} are subjected to LSA. While the main content of this paper deals with the Mood corpus, we use Genres to balance our data sampling in Section \ref{sec:gt}, and the other sets for comparison of different concepts in Section \ref{sec:hopkins}.

\subsection{Multidimensional Scaling}\label{sub:mds}

Past research in emotion modeling, reviewed above, suggests two to three underlying dimensions of emotions, which indicates that very concise representation of the mood data at hand would successfully explain most of its variance. Therefore, we develop further processing steps to produce a semantic space of moods congruent with the dimensional emotion model. Genres, Locales, Instruments and Opinions were subjected to the same procedures to allow comparative analysis described in Section \ref{sec:hopkins}.

We applied non-metric Multidimensional Scaling (MDS) \cite{Kruskal1964} according to Kruskal's Stress-1 criterion into three dimensions on the term dissimilarities produced by SVD with different rank $k$-values. MDS is a set of mathematical techniques for exploring dissimilarity data by representing objects geometrically in a space of a desired dimensionality, where the distances between objects in the obtained space approximate a monotonic transformation of the corresponding dissimilarities in the original data. When used with a low number of dimensions, MDS allows for concise representation of data, which is why it is a typical tool for data visualization. In particular, \cite{Becavin2011} showed with several high-dimensional biochemical data sets that the combination of SVD followed by MDS is more efficient at dimension reduction than either technique alone.

\begin{figure}[!t]
\subfigure{\includegraphics[trim = 1mm 0mm 0mm 0mm, clip, width=\columnwidth]{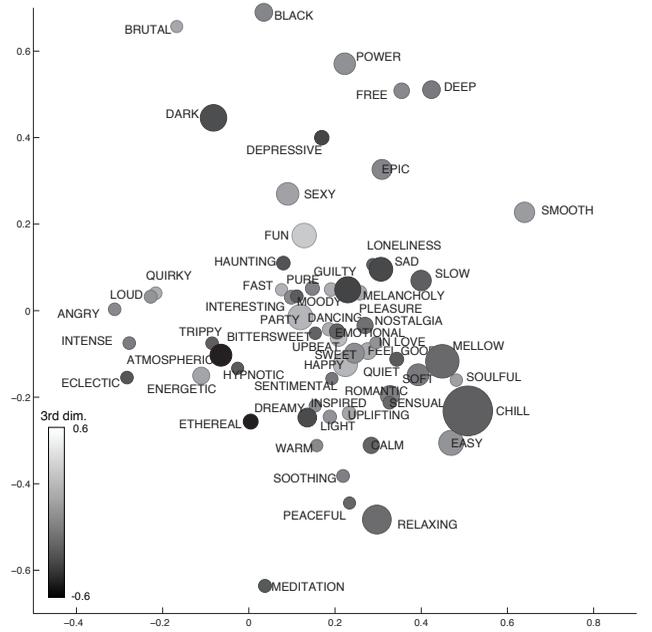}}%
\vspace{0.1cm}
\subfigure{\includegraphics[trim = 0mm 0mm 0mm 0mm, clip, width=\columnwidth]{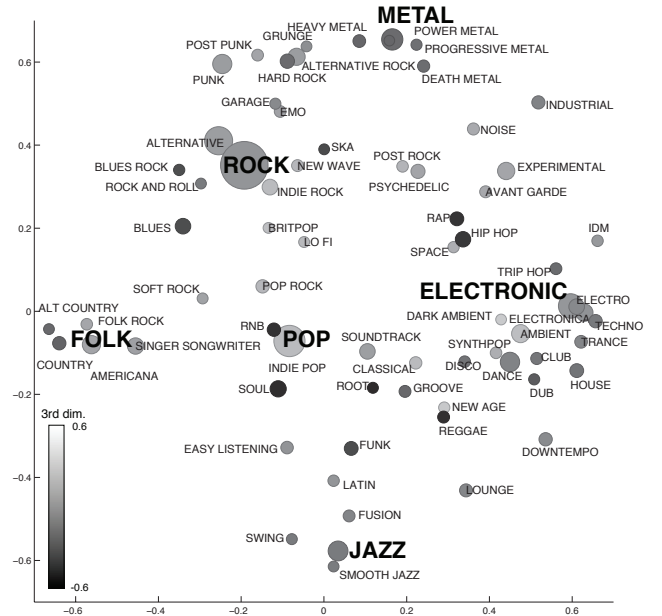}}%
\caption{MDS configurations ($k=16$) of (a) mood and (b) genre terms in three dimensions (bubble size = prevalence, where prevalence $\ge$ 4,000 and 10,000 for (a) and (b)). Six highlighted genres refer to listening experiment (see Section \ref{sec:gt}).}
\label{fig:mds_spaces}
\end{figure}

The resulting mood and genre term configurations with $k=16$ are shown in Fig. \ref{fig:mds_spaces}. The stress $\phi_k$, indicating the goodness-of-fit varied between ($\phi_4=0.02, \phi_{256}=0.29$) depending on the rank $k$. Similar values were obtained for both moods and genres.

To represent a track in the MDS term space, we applied projection based on the positions of the associated terms. Given an MDS term configuration $y_i = (y_{i1},y_{i2},y_{i3})$, $i \in (1,..., |w|)$, position of a track represented by a sparse term vector $q$ is computed by the center-of-mass:
\begin{equation}
\hat t=\frac{\Sigma_i q_{i} y_{i}}{\Sigma_i q_{i}}.
\label{eq:centerofmass}
\end{equation}
For example, the position of a track associated to ``happy'', with no other terms assigned, coincides with the position of the term. On the other hand, a track with ``happy'' and ``atmospheric'' is positioned along the segment \textit{happy--atmospheric}. In general, tracks are located in the MDS space within a \textit{convex polyhedron} with vertices defined by positions of the associated terms.

\subsection{Mood Structures Emerging from the Semantic Data}\label{sec:hopkins}

Because of the different views on how to treat mood-related data, whether as categories or dimensions, we used semantic information of music tracks obtained by the MDS analysis to gain evidence on this issue. If tracks in the MDS space would have clear cluster structure, we should choose the categorical representation; whereas, if tracks would scatter somewhat evenly across the space, continuous description of moods would be appropriate.

Hopkins' index \cite{Hopkins1954} can be used to estimate the degree of clusterability of multidimensional data. It is based on the hypothesis that the clustering tendency of a set of objects is directly reflected in a degree of non-uniformity in their distribution. Non-uniformity is estimated by comparing the sum of nearest-neighbor distances $R_j$ within a set of real objects to the sum of distances $A_j$ between artificial objects and their nearest real neighbors:
\begin{equation}
  H = \frac{\sum{A_j}}{\sum{A_j} + \sum{R_j}}.
  \label{eq:hopkins}
\end{equation}
Following an extension by \cite{lawson1990}, artificial objects are sampled from univariate distributions that match those of the real objects. Value $H \approx 0.50$ indicates uniform structure ($\sum R_j \approx \sum A_j$), whereas $H\approx 1.0$ indicates perfect clusterability. In particular, the value $H=0.75$ indicates that null hypothesis of uniform structure can be rejected at 90\% confidence level.

\begin{figure}[tp]
\centering
\includegraphics[trim= 4mm 4mm 9mm 7mm, clip, width=\columnwidth]{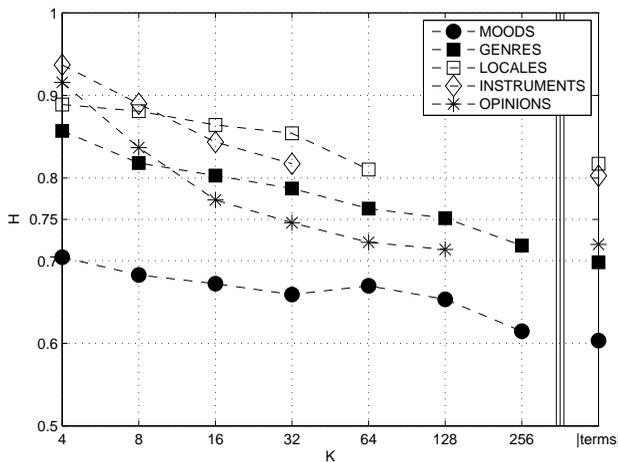}
\caption{The mean Hopkins' indices for each vocabulary-related corpus across various $k$ ($sd\leq.0067~\forall~k$).}
\label{fig:hopkins}
\end{figure}

For each corpus, we computed Hopkins' index for the track positions in the MDS spaces (see Eq. \ref{eq:centerofmass}) obtained with ranks $k=(4,8,16,...,256)$ and $k=|terms|$. The latter corresponds to computing MDS without LSA, i.e. based on term distances in the original TF-IDF matrices. Preliminary analyses indicated that Hopkins' index is affected by the number of terms associated to each track. Since the characteristics of the vocabulary-related corpora differed in this respect, we randomly sampled for each corpus a subset of $4088$ tracks with exponential terms-per-track distribution ($2048+1024+512+...+8$ tracks associated to $2,3,4,...,10$ terms, respectively) and computed $H$ for the subset. The results shown in Fig. \ref{fig:hopkins} are computed as an average of ten separate runs of this process.

The results showed that Hopkins' indices for Moods remained at the range of $0.6 < H < 0.7$, which means that track positions are uniformly distributed across the mood space. This suggests that the optimal representation of Moods is continuous rather than categorical. Comparison to other corpora supports this view, as mood values remain at a clearly lower level than those of any other set. Genre-related values indicated that genre-data is fairly clusterable ($H>.75$, when $k\leq128$), supporting the common practice of assigning songs categorically into genres. Furthermore, semantic spaces of Instruments and Locales had the clearest cluster structure. This is in line with the intuition that music tracks can, in general, be characterized with distinct instruments or instrumentations and geographical locations. Clusterability of data related to Opinions was in general at the same level as that of Genres. However, Opinions yielded particularly high values of $H$ with low $k$. We consider this high dependence on $k$ as an artefact caused by ill-conditioned distances between Opinion terms: almost all of the most prevalent terms were highly positive (``favorite'', ``killer'', ``amazing'', ``awesome'', etc.), and the computed distances between these terms may not reflect any true semantic relationships.

In summary, the results support the use of the dimensional representation of mood information of music tracks. In the next section we develop further processing steps to comply with this finding.

\section{Affective Circumplex Transformation}\label{sec:act}

Typical MDS operations, described above, may not be adequate to characterize moods, since the dimensions obtained do not explicitly represent the dimensional models of emotion. We therefore propose a novel technique called \textit{Affective Circumplex Transformation} (ACT) influenced by Russell's affective circumplex model of emotions \cite{Russell1980} to conceptualize the dimensions of the MDS mood spaces. First, reference positions for mood terms on the Valence-Arousal (VA) space are obtained from past research on emotion modeling. Then, the MDS space is linearly transformed to conform to the reference. Finally, explicit mood information of music tracks is computed by projecting those onto the transformed space.

\subsection{ACT of Mood Term Space}

Reference locations for a total of 101 unique mood terms on the VA space were extracted from Russell's~\cite[p. 1167]{Russell1980} and Scherer's~\cite[p. 54]{Scherer1984} studies. In the case of seven overlapping mood terms between the two studies, Scherer's term positions were chosen since they are scattered on a larger part of the plane and thus may provide more information. Furthermore, the model by \cite{Thayer1989} was projected on the space diagonally against the negative valence and positive arousal to obtain explicit representation of the tension dimension.

Three-dimensional MDS spaces were conformed to the extracted VA space by first identifying the corresponding mood terms in the semantic data. Identification of mood terms resulted in a set of 47 mood terms out of the 101 candidates. The fact that less than half of the mood terms used in the past studies exist in the semantic mood data may indicate the difference between affect terms used to describe everyday experiences in general versus terms used in the context of the aesthetic experience.

Transformation of the MDS space to optimally conform to the VA reference was determined by classical Procrustes analysis \cite{Gower2004}, using sum of squared errors as goodness-of-fit. Given the MDS configuration $y_{\hat i} = (y_{\hat i1},y_{\hat i2},y_{\hat i2})$ and VA reference $x_{\hat i} = (x_{\hat i1},x_{\hat i2})$ for mood terms $\hat i$ matched between the two, Procrustes transformation gives $\hat x_{\hat i} = B y_{\hat i} T + C$, where $B$ is an isotropic scaling component, $T$ is an orthogonal rotation and reflection component, and $C$ is a translation component. $B$, $T$, and $C$ minimize the goodness-of-fit measure $X^2 = \Sigma_{\hat i} (x_{\hat i} - \hat x_{\hat i})^2$. Based on the components, configuration $\hat x_{i}$ including all mood terms can be obtained by
\begin{equation}
\hat x_{i} = B y_{i} T + C.
\end{equation}
Procrustes retains the relative distances between objects since it allows only translation, reflection, orthogonal rotation and isotropic scaling. Therefore, the relative configuration of the terms in the original MDS space is not affected. Changing the rank parameter in SVD had no significant effect on the goodness-of-fit of the Procrustes transformation. The criterion varied between $0.75 < X^2 < 0.79$.

A peculiarity of ACT is in conforming the three-dimensional MDS space to two-dimensional reference. The transformation is thus provided with an additional degree of freedom, producing two explicitly labeled dimensions and a third residual dimension. Using three dimensions in the MDS space is based on the unresolved debate of whether the underlying emotion space is actually two- or three-dimensional (see Section \ref{sec:moodspsych}).

\begin{figure}[!tp]
\centering
\includegraphics[width=\columnwidth]{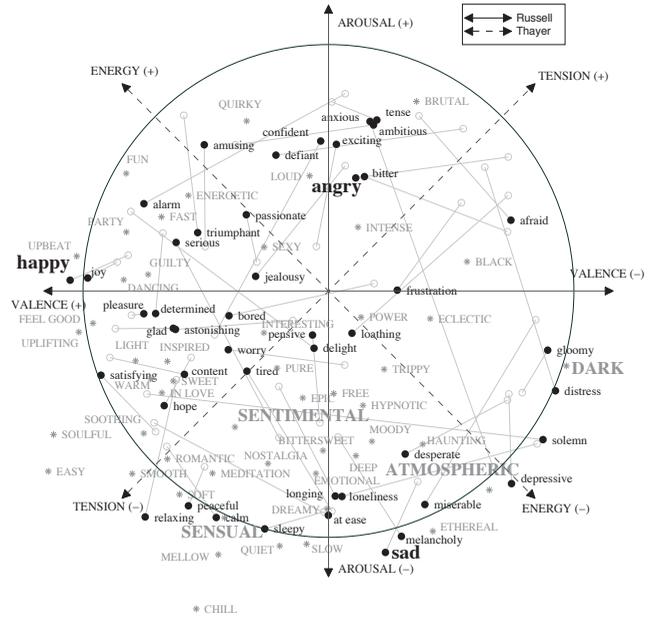}
\caption{Mood space obtained by ACT based on \textit{rank $16$ semantic space}. Mood term positions are shown in black dots and the reference positions in grey circles. Mood terms having no reference, but that are associated to at least 4,000 tracks are shown in grey stars. Highlighted terms relate to the seven scales rated in listening experiment (see Section \ref{sec:gt}).}
\label{fig:russell}
\end{figure}

Fig. \ref{fig:russell} shows the transformed mood term configuration based on SVD with rank $16$, also indicating Russell's dimensions of Arousal and Valence, and Thayer's dimensions of Energy and Tension. VA-reference and the transformed term positions correspond well, in general, as they are located roughly at the same area of the space. For example, positions of terms ``happy'', ``joy'', ``sad'', ``tense'' and ``peaceful'' have only minor discrepancy between the reference. Moreover, dimension labels and the dimensions implied by the mood term organization correspond as well and the positions of popular mood terms not used as reference for the transformation make sense in general. For example, ``fun'', ``party'' and ``upbeat'' all have positive valence and arousal, ``dark'' has negative valence and negative arousal, whereas ``brutal'' has negative valence and positive arousal.

However, certain terms such as ``solemn'', ``delight'', ``distress'' and ``anxious'' show larger discrepancy, and the terms ``atmospheric'' and ``ethereal'', which could intuitively be considered as neutral or even positive, both have negative valence. The cause of these inconsistencies could again be traced back to the difference between aesthetic and everyday affective experience, but could also be due to the subjectivity of mood-related associations in music listening. For example, a solemn or atmospheric track that one enjoys may be regarded as depressing by another. This multi-faceted aspect of music listening is discussed in \cite{juslin2011emotional}.

\subsection{Mood Prediction of Tracks with ACT}

In this section we describe how ACT can be used to predict the prevalence of moods in novel tracks that have been associated to one or more mood terms. The prediction performance of ACT related to mood dimensions and individual mood terms is evaluated later in Section \ref{sec:results} with listener ratings of the perceived mood in a separate test set.

Projection in Eq. \ref{eq:centerofmass} can be used in itself to estimate the valence and arousal of a track -- the estimations are represented explicitly by the dimensions in the projection. However, in order to estimate the prevalence of a certain mood term in a track, another projection is needed.

We assigned continuous mood term-specific weight for a track by projecting the track position given by Eq. \ref{eq:centerofmass} in the MDS space along the direction determined by the term. A track with position $\hat t = (\hat t_{1}, \hat t_{2}, \hat t_{3})$ in the transformed mood space was projected according to the direction of a mood term with position $\hat x_i = (\hat x_{i1}, \hat x_{i2}, \hat x_{i3})$ by
\begin{equation}
  P_{i} = \frac{\hat x_i}{|\hat x_i|} \cdot \hat t,
  \label{eq:projection}
\end{equation}
where $|\cdot|$ denotes the $l^2$-norm. To obtain an estimate for Tension we projected the track along the direction $(-1,1,0)$ (note the inverted valence axis according to a convention used in emotion modeling). 

\section{Baseline techniques for Mood Prediction}\label{sec:baseline}

We compared the performance of ACT in predicting moods of tracks to several techniques based on low-rank matrix approximation. For a track $t$ represented by a sparse term vector $q$, we computed rank $k$ mood weight related to term $w_i$ with SVD, NMF and PLSA. All of the techniques involve computing the low-rank approximation of the TF-IDF matrix, transforming an unknown track $t$ to the VSM by Eq. \ref{eq:tfidf} and folding it into the low-rank semantic space, and approximating the weight of a mood $w_i$ related to the track. In addition to the low-rank approximation of the VSM, we used the original sparse VSM representation $q$ as a baseline as well.

\subsection{Singular Value Decomposition}

Track represented with a sparse term vector $q = (q_{1}, q_{2}, ..., q_{|w|}, )$ is first folded in to the rank $k$ space obtained with SVD by:
\begin{equation}\label{eq:foldingin}
\hat{q}^k = (S^k)^{-1} (U^k)^Tq.
\end{equation}

The weight $N^k_{i}$ related to the track and a mood term $w_i$ is then computed by
\begin{equation}
N^k_{i} = U^k_iS^k(\hat q^k)^T.
\end{equation}

\subsection{Nonnegative Matrix Factorization}

NMF \cite{Lee2001} is a method proposed for low-rank approximation of a term-document matrix. The method distinguishes from SVD by its use of nonnegative constraints to learn parts-based representation of object semantics. Given a nonnegative TF-IDF matrix $N\subset \mathbb{R}^{C\times D}$ and a desired rank parameter $k$, NMF constructs nonnegative matrices $W^k \subset \mathbb{R}^{C\times k}$ containing $k$ basis components and $H^k \subset \mathbb{R}^{k\times D}$ such that $N\approx W^kH^k$. This is done by optimizing
\begin{equation}\label{eq:nmf}
\min_{W^k,H^k}f(W^k,H^k)=\frac{1}{2}|| N - W^kH^k  ||^2_F,~s.t.~W^k, H^k>0,
\end{equation}
where $F$ denotes the Frobenius norm. We solve the optimization problem using multiplicative updating rules in an iterative manner \cite{Lee2001}. The $i$th row of $W$ can be interpreted as containing $k$ ``importance" weights a mood term $w_i$ has in each basis component. Similarly, the $j$th column of $H$ can be regarded as containing $k$ corresponding weighting coefficients for track $t_j$.

Folding in a new track represented by vector $q$ to obtain $\hat{q}^k$ is achieved by solving an optimization problem by keeping $H^k$ fixed:
\begin{equation}
\min_{\hat{q}^k}f(\hat{q}^k,H^k)=\frac{1}{2}|| q - \hat{q}^kH^k  ||^2_F,~s.t.~\hat{q}^k>0.
 \end{equation}
Finally, to estimate the weight $N^k_{i}$ related to track $t$ and mood term $w_i$, we compute

\begin{equation}
N^k_{i} = W^k_i\hat q^k.
\end{equation}

\subsection{Probabilistic Latent Semantic Analysis}

In the core of PLSA \cite{Hofmann2001}, is the statistical \textit{aspect model}, a latent variable model for general co-occurrence data. Aspect model associates an unobserved class variable $z \in Z = (z_1, ..., z_k)$ with each occurrence of a term $w_i$ in a track $t_j$.

PLSA states that the probability $P(t_j,w_i)$ that term $w_i$ is associated with a track $t_j$ can be expressed as a joint probability model using latent class variable $z$:
\begin{equation}
P(t_j,w_i)=P(t_j)P(w_i|t_j)=P(t_j)\sum_{z\in Z}P(w_i|z)P(z|t_j),
\end{equation}
where $P(t)$ is the probability of a track $t_j$, $P(z|t_j)$ is the probability of a latent class $z$ in track $t_j$, and $P(w_i|z)$ is the probability of a term $w_i$ in the latent class. The model is fitted to the collection of tracks by maximizing log-likelihood function
\begin{equation}
L=\sum_t\sum_w N_{i,j}logP(t_j,w_i),
\end{equation}
where $N_{i,j}$ is the nonnegative TF-IDF matrix. The procedure for fitting the model to training data is the Expectation Maximization (EM) algorithm \cite{Hofmann2001}. To estimate the probability $P(q,w_i)$ of a mood term $w_i$ for a new track represented by term weights, we first fold in the track using EM, keeping the parameters of $P(w_i|z)$ fixed and then calculate weights $P(z|q)$. The mood weight for the track is finally computed by
\begin{equation}
P(q,w_i)=P(w_i|z)P(z|q).
\end{equation}
\subsection{Predicting the Mood Dimensions}

Since all baseline techniques predict mood primarily according to explicit mood terms, the techniques must be optimised to achieve mood dimension predictions comparable to ACT. We considered that a mood term representative of a mood dimension would yield the highest predictive performance for the corresponding dimension. We assessed the representativeness of the mood terms by computing the angle between each mood dimension and mood term location in the ACT configurations with $k \in [4,8,16,...,256]$, and limited the choice to terms associated to at least 10\% of all tracks in the corpus. This yielded the following terms, indicating the number of track associations and the maximum angle across $k$ between the term position in the ACT configurations and the corresponding dimension: ``happy'' for Valence ($n=28,982$, $\alpha_k \leq 9.29^{\circ}$), ``melancholy'' for Arousal ($n=31,957$, $\alpha_k \leq 5.11^{\circ}$) and ``mellow'' for Tension ($n=46,815$, $\alpha_k \leq 4.48^{\circ}$)

\section{Ground-Truth Data of Moods in Music}\label{sec:gt}

We evaluated the performance of ACT and the baseline techniques by comparing the estimates produced by these methods to listener ratings of the perceived moods in music tracks. Participants listened to short music clips (15s) and rated their perception of moods expressed by music in terms of ten scales. The test set of tracks was retrieved from the Last.fm in a random fashion, balancing the sampling to cover semantic genre and mood spaces. This section describes the ground-truth collection process in detail\footnote{Ground-truth and semantic mood data are publicly available at \url{http://hdl.handle.net/1902.1/21618}.}.

\subsection{Choosing Moods and Genres as Focus}

To systematically cover the concurrent characterizations of moods in music, ratings were done for both the dimensional mood model and individual mood terms. All ratings were given in nine-step Likert-scales to capture the continuous nature of mood uncovered in Section \ref{sec:hopkins}. We used bipolar and unipolar scales for the mood dimensions and terms, respectively.

For dimensional model we used three scales: Valence, Arousal and Tension, later denoted as VAT; whereas for the mood term representation we used seven scales: Atmospheric, Happy, Dark, Sad, Angry, Sensual and Sentimental. The choice was based on several criteria: {\em i)} to cover the semantic space as well as the basic emotion model; {\em ii)} to use representative terms as implied by high prevalence in the data (``sentimental'' used 4,957 times -- ``dark'' 33,079 times); and {\em iii)} to comply with research in the affect prevalence and applicability in music \cite{Juslin2004,Zentner2008,juslin2011emotional}.

Six popular and distinct genres according to the Last.fm track collection (see Fig. \ref{fig:mds_spaces} (b)) -- Rock, Pop, Electronic, Metal, Jazz and Folk -- were chosen as the focus of the study to retain a wide variance in the stylistic characteristics of popular music.

\subsection{Sampling of Tracks}

We fetched a set of $600$ tracks from Last.fm, separate to the mood track corpus used in the semantic modeling, to be rated in the listening experiment. To obtain a track collection that allows multifaceted comparison between tag information and the ratings, we utilized balanced random sampling of tracks based on: {\em i)} mood coverage -- reciprocal of the track density in the rank $16$-based MDS mood space; and {\em ii)} genre coverage -- closeness of track positions in the MDS genre space to one of the six chosen genre terms. Moreover, quality and variability of semantic information in the data was ensured by: {\em i)} favoring tracks associated to many mood tags; {\em ii)} favoring tracks with many listeners according to statistics provided by Last.fm; and {\em iii)} choosing no more than one track from each artist.

\begin{figure}[!t]
\includegraphics[trim=15mm 0mm 16mm 1mm, clip, width=\columnwidth]{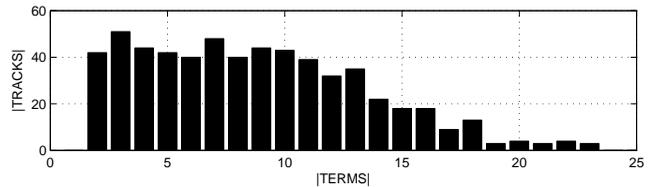}%
\caption{Distribution of the number of mood terms associated to tracks in the test set.}
\label{fig:tagsums600}
\end{figure}

Tracks in the resulting test set are associated with $8.7$ mood terms on average, which is a higher number than that of the larger mood set due to sampling according to the number of associated mood terms. The details of the term occurrences are shown in Fig. \ref{fig:tagsums600}. The number of times each mood term related to the chosen scales appear in the set are: $90$ (Atmospheric), $137$ (Happy), $109$ (Dark), $166$ (Sad), $28$ (Angry), $43$ (Sensual) and $52$ (Sentimental). For genres, the corresponding figures are: $422$ (\textit{rock}), $353$ (\textit{pop}), $149$ (\textit{electronic}), $139$ (\textit{metal}), $147$ (\textit{jazz}) and $144$ (\textit{folk}). Considering the high frequency of genres such as \textit{rock} and \textit{pop}, these genres have naturally wider representation in the set -- a track in the \textit{electronic} genre has likely been tagged with \textit{pop}, for instance.

\subsection{Listening Experiment}

An online interface was used to allow participants to login on their own computers and save their ratings on a server in real time. At each session, tracks were presented in a randomized order. Participants were allowed to rate as many or as few songs as they liked. However, to encourage the rating of many tracks, the task was rewarded by Spotify\footnote{\url{http://www.spotify.com/}} and Amazon\footnote{\url{http://www.amazon.co.uk/}} gift cards proportional to the amount of tracks rated.

The task was to rate $15$ second clips of the tracks in terms of the perceived moods \textit{expressed} by music, rather than moods \textit{induced} by music. VAT scales were presented with bipolar mood term labels: ``negative''/``positive'', ``calm''/``energetic'' and ``relaxed''/``tense'', respectively. In addition to mood, participants rated their personal liking of the tracks, and in half of the cases, genre representativeness. In this paper, however, we utilize only the mood ratings.

We based the sampling of each song on the audio previews on Last.fm service, arguing that, since the previews are track summarizations sampled for marketing purposes, consisting of the most prolific section, they are fairly representative of the full tracks. The previews typically consist of a build-up and part of chorus, starting either at $30$ or $60$ seconds into the beginning. While some studies have highlighted the difference between clip- and track-level content \cite{Mandel2011}, it has been argued that using short clips lessens the burden of human evaluation and reduces problems in annotation caused by time variation of moods \cite{Hu:08}.

A total of $59$ participants, mostly Finnish university students (mean age $25.8$ years, SD $=5.1$ years, $37$ females), took part in the experiment. Musical expertise of the participants spanned from listeners ($N=23$), to musicians ($N=28$) and trained professionals ($N=8$). Each participant rated $297$ clips on average, and $22$ participants rated all $600$ clips. Cronbach's alpha for mood scales vary between $0.84$ (sentimental) and $0.92$ (arousal), which indicates high internal consistency \cite{Nunnally1978}. Such high agreement among the participants gives support for (a) using all participants in further analysis, and (b) representing each song by single value on each mood scale, computed as the average across participants.

Spearman's rho correlations ($r_s$) between mood ratings in different scales, presented in Table \ref{table:moodcorr}, showed no correlation between valence and arousal, which supports treating these moods as separate dimensions. On the other hand, tension is highly correlated with arousal and negative valence, which in turn supports projecting tension diagonally against these dimensions. Ratings of all $7$ mood terms are highly related to valence (happiness, darkness), arousal (atmospheric, sentimental), or a combination of these (sad, angry, sensual). This extends previous findings about high congruence between term-based and dimensional emotion models in emotion ratings of film soundtracks \cite{Eerola2011} to a large variety of tracks in popular music genres.

\begin{table}
\caption{Correlations ($r_s$) between mood ratings.}
\label{table:moodcorr}
\centering
\begin{tabular}{llll}
&Valence&Arousal&Tension\\
\hline
Valence&&$-.073^{}$&$-.639^{***}$\\
Arousal&&&\hspace{2.4mm}$.697^{***}$\\
\cline{2-4}
Atmospheric&\hspace{2.4mm}$.180^{***}$&$-.901^{***}$&$-.687^{***}$\\
Happy&\hspace{2.4mm}$.940^{***}$&\hspace{2.4mm}$.114^{**}$&$-.478^{***}$\\
Dark&$-.940^{***}$&\hspace{2.4mm}$.059$&\hspace{2.4mm}$.640^{***}$\\
Sad&$-.413^{***}$&$-.662^{***}$&$-.253^{***}$\\
Angry&$-.687^{***}$&\hspace{2.4mm}$.633^{***}$&~~$.876^{***}$\\
Sensual&\hspace{2.4mm}$.320^{***}$&$-.733^{***}$&$-.688^{***}$\\
Sentimental&\hspace{2.4mm}$.114^{**}$&$-.722^{***}$&$-.621^{***}$\\
\hline
\multicolumn{3}{l}{\scalebox{.8}{Note: $^{*}p<.05$; $^{**}p<.01$; $^{***}p<.001$, $df=599$.}}
\end{tabular}
\end{table}

\section{Results and Discussion}\label{sec:results}

\begin{figure*}[!t]
\centerline{
\subfigure{\includegraphics[trim=50mm 123mm 352mm 11mm, clip, height=4.5cm]{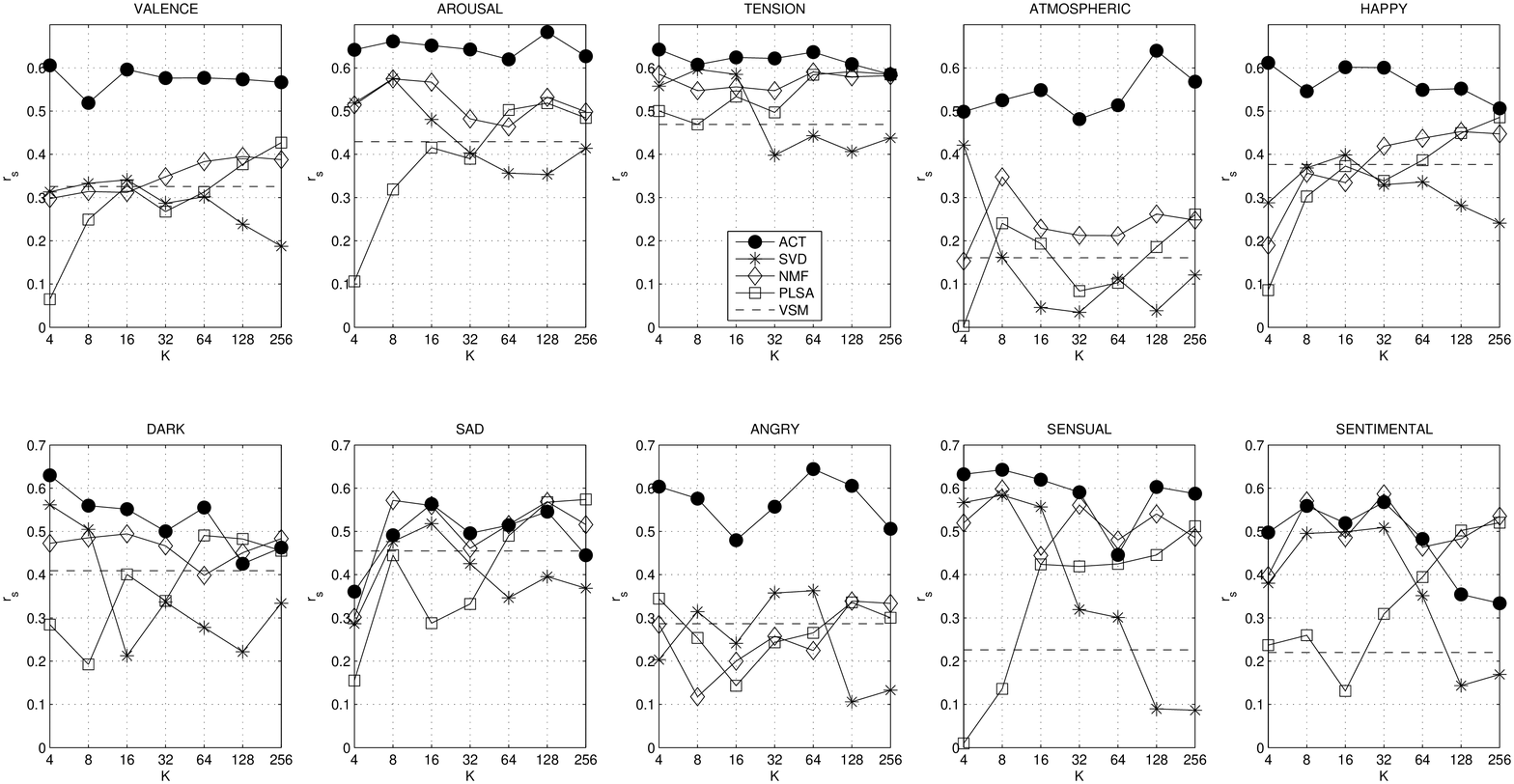}}
\hfill
\subfigure{\includegraphics[trim=133mm 123mm 274mm 11mm, clip, height=4.5cm]{correlations_all}}
\hfill
\subfigure{\includegraphics[trim=211mm 123mm 196mm 11mm, clip, height=4.5cm]{correlations_all}}
\hfill
\subfigure{\includegraphics[trim=288mm 123mm 119mm 11mm, clip, height=4.5cm]{correlations_all}}
\hfill
\subfigure{\includegraphics[trim=365mm 123mm 42mm 11mm, clip, height=4.5cm]{correlations_all}}
}
\centerline{
\subfigure{\includegraphics[trim=50mm 16mm 352mm 118mm, clip, height=4.5cm]{correlations_all}}
\hfill
\subfigure{\includegraphics[trim=133mm 16mm 274mm 118mm, clip, height=4.5cm]{correlations_all}}
\hfill
\subfigure{\includegraphics[trim=211mm 16mm 196mm 118mm, clip, height=4.5cm]{correlations_all}}
\hfill
\subfigure{\includegraphics[trim=288mm 16mm 119mm 118mm, clip, height=4.5cm]{correlations_all}}
\hfill
\subfigure{\includegraphics[trim=365mm 16mm 42mm 118mm, clip, height=4.5cm]{correlations_all}}
}
\caption{Prediction rates ($r_s$) of listener ratings in (a -- c) VAT scales and (d -- j) mood term scales.}
\label{fig:correlations}
\end{figure*}

We compared the prediction rates of ACT with various rank values $k \in (4,8,16,...,256)$ to those of the baseline techniques SVD, NMF, PLSA and VSM. All prediction rates were computed by correlating the estimates with the listener ratings of moods, using Spearman's rank correlation ($r_s$). Fig. \ref{fig:correlations} shows the results in detail with different rank $k$ values, while Table \ref{table:performance} summarizes the results into the average performance across $k$, assessing also the significance of the performance differences between ACT and the baseline techniques. Section \ref{sec:act_alt} (Table \ref{table:performance}: ACT alt.) provides results obtained with alternative configurations of ACT. Finally, Section \ref{sec:termdensity} assesses the performance of ACT as a factor of the number of terms applied to tracks in the test set.

\begin{table*}
\caption{Comparison of the performances of ACT, baseline techniques, and alternative ACT implementations (\textit{Mdn} = median across $k$). Significances of the performance differences were computed by Wilcoxon rank sum test for equal medians between ACT and SVD, NMF and PLSA, and Wilcoxon signed rank test for median $r_s$ between ACT and VSM, SVD only, and MDS only.}
\label{table:performance}
\centering
\begin{tabular}{llll||llll ||| ll}
\multicolumn{1}{c}{}&\multicolumn{3}{c}{\underline{ACT}}&\multicolumn{4}{c}{\underline{BASELINE}}&\multicolumn{2}{c}{\underline{ACT alt.}}\\
\multicolumn{1}{c}{}&\multicolumn{3}{c}{}&SVD&NMF&PLSA&\multicolumn{1}{l}{VSM}&SVD-only&MDS-only\\
\cline{2-10}
\multicolumn{1}{c}{}&$r_s$ \textit{(Mdn)}&$r_s$ \textit{(min)}&$r_s$ \textit{(max)}&$r_s$ \textit{(Mdn)}&$r_s$ \textit{(Mdn)}&$r_s$ \textit{(Mdn)}&$r_s$&$r_s$&$r_s$\\
\hline
Valence&$\textbf{.576}^{}$&$.519^{}$&$.606^{}$&$.302^{***}$&$.348^{***}$&$.313^{***}$&$.326^{*}$&$.475^{*}$&$.558^{}$\\
Arousal&$\textbf{.643}^{}$&$.620^{}$&$.683^{}$&$.414^{***}$&$.514^{***}$&$.416^{***}$&$.429^{*}$&$.373^{*}$&$\textbf{.643}^{}$\\
Tension&$\textbf{.622}^{}$&$.585^{}$&$.642^{}$&$.443^{**}$&$.579^{**}$&$.534^{**}$&$.469^{*}$&$.591^{*}$&$.596^{*}$\\
\cline{2-10}
Atmospheric&$\textbf{.525}^{}$&$.482^{}$&$.640^{}$&$.112^{***}$&$.229^{***}$&$.186^{***}$&$.161^{*}$&$.247^{*}$&$\textbf{.581}^{}$\\
Happy&$\textbf{.552}^{}$&$.506^{}$&$.612^{}$&$.330^{***}$&$.419^{***}$&$.373^{***}$&$.376^{*}$&$.279^{*}$&$.455^{*}$\\
Dark&$\textbf{.552}^{}$&$.425^{}$&$.630^{}$&$.334^{}$&$.472^{}$&$.401^{*}$&$.409^{*}$&$\textbf{.595}^{*}$&$.239^{*}$\\
Sad&$.496^{}$&$.361^{}$&$.563^{}$&$.396^{}$&$\textbf{.516}^{}$&$.445^{}$&$.455^{}$&$.328^{*}$&$.469^{}$\\
Angry&$\textbf{.576}^{}$&$.480^{}$&$.644^{}$&$.241^{***}$&$.258^{***}$&$.265^{***}$&$.286^{*}$&$-.131^{*}$&$.432^{*}$\\
Sensual&$\textbf{.603}^{}$&$.446^{}$&$.643^{}$&$.319^{**}$&$.520^{*}$&$.424^{**}$&$.226^{*}$&$.589^{}$&$.542^{}$\\
Sentimental&$\textbf{.498}^{}$&$.334^{}$&$.568^{}$&$.380^{}$&$.486^{}$&$.309^{}$&$.220^{*}$&$.420^{}$&$.356^{}$\\
\hline
\multicolumn{3}{l}{}&&\multicolumn{4}{l}{\scalebox{.8}{Note: $^{*}p<.05$; $^{**}p<.01$; $^{***}p<.001$, $df=6$.}}\\
\end{tabular}
\end{table*}

\subsection{Performance for VAT Dimensions}
\label{sub:model_performance}

Fig. \ref{fig:correlations} shows that ACT yielded the highest performance for all VAT scales, outperforming the baseline techniques consistently across $k$. For Valence the median performance of ACT was $r_s=.576$, varying between $.519<r_s<.606$. The performance was slightly higher for Arousal (Mdn $r_s=.643$, $.620<r_s<.683$) and Tension (Mdn $r_s=.622$, $.585<r_s<.642$). Performance difference to the baseline techniques was significant for all scales -- NMF gave the highest median performances ($r_s=.348, .514, .579$), while SVD performed the worst ($r_s=.302, .414, .443$) at predicting Valence, Arousal and Tension, respectively. VSM yielded performance levels comparable to the baseline methods, outperforming SVD for all three scales, and PLSA for Valence and Arousal. However, devising baseline techniques to infer predictions for VAT scales from highly prevalent mood terms possibly benefits VSM more than the other techniques. While SVD, NMF and PLSA utilize the semantic relationships with other terms in making predictions, VSM predictions rely solely on the individual terms. The chosen mood terms are popular also within the test set ($n=137,189,227$ for ``happy'', ``melancholy'' and ``mellow'', respectively).

The results also show that ACT is less sensitive to the value of $k$ than SVD, NMF and PLSA. While ACT performance varied by $\Delta r_s \leq .087$, SVD ($\Delta r_s \leq .222$) and PLSA ($\Delta r_s \leq .412$) were clearly more inconsistent. For Valence and Arousal, PLSA yielded particularly low performance with $k<16$. NMF was more robust than other baseline techniques against $k$ as shown by the performance differences of $\Delta r_s \leq .112$.

The high prediction rate of Arousal compared to that of Valence bears similarity to the results from prediction of affect dimensions from the musical features across different genres of music \cite{eerola2011c}. This was also highlighted by an analysis of ACT prediction rates at the genre-level. The median $r_s$ across $k$ for subsets of the test tracks associated to different main genres was consistently high for Arousal regardless of genre ($.585<r_s<.701$), whereas for Valence the rates spanned $r_s=.390$ (Jazz) and $r_s=.614$ (Metal).

In summary, the results suggest that conventional techniques of semantic analysis are inadequate at reliably inferring mood predictions congruent with the dimensional model of emotions, whereas ACT yields consistently high performance at this task.

\subsection{Performance for Individual Mood Terms}
\label{sub:subsection_name}

Since the rated mood term scales relate to the mood term associations explicitly represented in the test set, comparison between ACT and the baseline techniques is more direct than with VAT dimensions. Still, the same patterns in the performances were prevalent. ACT, again, clearly gave the highest overall performance, while NMF was the most successful baseline method. NMF outperformed ACT only at predicting Sad, but this difference was not, however, statistically significant.

In general, median performances of ACT were lower for the individual mood scales than for VAT dimensions, ranging from $r_s=.496$ (Sad) to $r_s=.603$ (Sensual). Performance difference between ACT and baseline techniques was the most notable for Atmospheric and Angry. While ACT yielded median performances $r_s=.525$ for the former scale and $r_s=.576$ for the latter, the most successful baseline techniques (NMF and VSM, respectively) produced only $r_s=.229$ and $r_s=.286$.

ACT performance was generally more sensitive to the value of $k$ for the individual mood terms than for the VAT dimensions. The performance range was smallest for Happy ($\Delta r_s=.105$, $.506 \leq r_s \leq .612$) and largest for Sentimental ($\Delta r_s=.234$, $.334 \leq r_s \leq .568$). However, the corresponding numbers were higher for all baseline techniques.

All in all, these results show that ACT is efficient at predicting the individual mood terms and gives consistent performance for mood terms (Atmospheric, Angry), which the baseline techniques fail at predicting. Together with the findings for VAT dimensions, this suggests that domain knowledge on moods can be utilized to great benefit in semantic computing.

\subsection{ACT with Alternative Implementations}\label{sec:act_alt}

While ACT clearly outperformed the baseline techniques at predicting the perceived mood, we carried out further comparative performance evaluation with ACT to assess the optimality of the technique. In particular, we were interested to find whether it is beneficial to implement ACT with dimension reduction in two stages, involving low-rank approximation with SVD and mood term configuration with MDS. For this evaluation we analyzed the performance of two models: {\em{a)}} SVD-only applying Procrustes directly on the SVD mood term configuration $u_i=U_i^kS^k$ ($k=3$) without the MDS stage; and {\em{b)}} MDS-only applying MDS on the cosine distances between mood terms computed from the raw TF-IDF matrix instead of the low-rank representation. In must be noted, however, that the latter model effectively corresponds to the original ACT with $k=|terms|$ but is computationally heavier than the original ACT when the TF-IDF matrix is large.

The results presented in Table \ref{table:performance} show that both ACT implementations yielded performance mostly comparable to that of the original ACT. The original ACT generally outperformed both alternative implementations. This difference was statistically significant in seven moods for SVD-only and in four moods for MDS-only. SVD-only outperformed the original ACT for Dark, whereas MDS-only yielded the highest performance for Arousal and Atmospheric. However, the performance differences for MDS-only were not statistically significant. The clearest difference was between ACT and SVD-only for Angry, where SVD-only failed to produce positive correlation.

The results suggest that mood prediction performance of ACT is significantly boosted by utilizing both SVD and MDS. 

\subsection{The Effect of Tag Sparsity on ACT Performance} \label{sec:termdensity}

\begin{table*}
\caption{Median performances ($r_s$) across $k$ obtained with ACT when the specified numbers of mood terms in average were associated to each track in the test set. \textit{\# Tracks} refers to the number of the fetched Last.fm tracks with at least \textit{\# Terms}.}
\label{table:termdensity}
\centering
\begin{tabular}{l cccccccc || c}
\multicolumn{1}{l}{\textit{\# Terms / Track}}&\textbf{\underline{1}}&\textbf{\underline{2}}&\textbf{\underline{3}}&\textbf{\underline{4}}&\textbf{\underline{5}}&\textbf{\underline{6}}&\textbf{\underline{7}}&\multicolumn{1}{c}{\textbf{\underline{8}}}&\multicolumn{1}{c}{\underline{\textbf{8.71 (Full)}}}\\
\multicolumn{1}{l}{\textit{\# Tracks}}&\textit{493,539}&\textit{259,593}&\textit{164,095}&\textit{114,582}&\textit{84,206}&\textit{64,018}&\textit{49,393}&\multicolumn{1}{l}{\textit{38,450}}&\\
\hline
Valence&$.445$&$.474$&$.498$&$.521$&$.535$&$.548$&$.558$&$.568$&$.576$\\
Arousal&$.492$&$.530$&$.560$&$.578$&$.600$&$.615$&$.627$&$.639$&$.643$\\
Tension&$.496$&$.535$&$.559$&$.576$&$.590$&$.598$&$.607$&$.617$&$.622$\\
\cline{2-10}
Atmospheric&$.375$&$.419$&$.445$&$.462$&$.477$&$.493$&$.509$&$.519$&$.525$\\
Happy&$.418$&$.454$&$.479$&$.497$&$.513$&$.525$&$.535$&$.543$&$.552$\\
Dark&$.413$&$.447$&$.471$&$.495$&$.512$&$.527$&$.539$&$.545$&$.552$\\
Sad&$.368$&$.387$&$.410$&$.429$&$.451$&$.471$&$.482$&$.491$&$.496$\\
Angry&$.490$&$.511$&$.525$&$.540$&$.546$&$.554$&$.562$&$.570$&$.576$\\
Sensual&$.475$&$.510$&$.535$&$.550$&$.567$&$.578$&$.586$&$.595$&$.603$\\
Sentimental&$.352$&$.382$&$.410$&$.428$&$.450$&$.463$&$.477$&$.489$&$.498$\\
\hline
\end{tabular}
\end{table*}

\begin{figure}[!t]
\centerline{
\subfigure{\includegraphics[trim=12mm 0mm 106mm 0mm, clip, height=4cm]{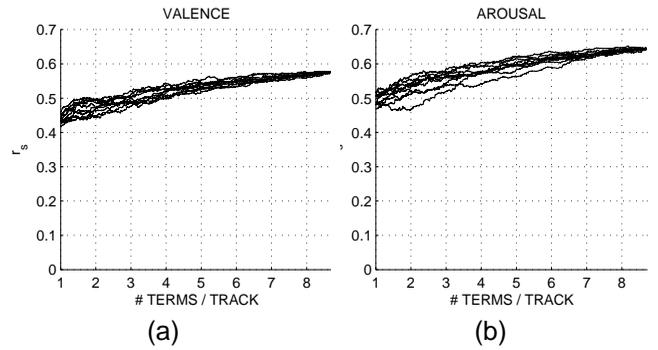}}
\hfill
\subfigure{\includegraphics[trim=104mm 0mm 18mm 0mm, clip, height=4cm]{termdensity_all}}
}
\caption{The relationship between the number of tags for each track and the ACT performance ($k=32$) for (a) Valence and (b) Arousal.}
\label{fig:termdensity}
\end{figure}

As noted in the introduction, social tag data is sparse, meaning that a typical document is associated to only a subset of all relevant tags. In the mood data fetched from Last.fm \textit{493,539} tracks are associated to at least one mood term, whereas only 38,450 tracks are associated to at least $8$ terms, which is approximately the average within the test set. If we consider only the level of data sparsity, we can assume that the performance presented above extends to approximately \textit{38,000} tracks. The question is, how high could prediction performance be expected for the larger set of almost \textit{500,000} tracks?

To study this question, we carried out systematic performance assessment with ACT as a factor of the number of mood terms associated to the test tracks. Starting with the full test set, we iteratively removed one term-track association at a time until all tracks were associated to only one term. The association to be removed was sampled in a weighted random manner at each iteration, weighting tracks directly proportional to the number of associated terms, and terms with lower raw within-track counts. We recorded ACT performance at each iteration, and calculated the mean performance across ten separate runs. The process can be seen as imitating user tagging, where a novel track in a system is most likely first applied with clearly descriptive tags. The results of the analysis are summarized in Table \ref{table:termdensity}, showing the median performance across $k$ obtained with $1, 2, 4, 6$ and $8$ terms associated to each track in average.

The results suggest that tag sparsity and prediction performance are in a strong linear positive relationship, supporting the assumption that tag sparsity primes the ACT prediction performance. This relationship held also at each of the ten separate runs (see Fig. \ref{fig:termdensity}). Depending on the mood, performance achieved with only one tag in each track was approximately $r_s=.433$ and varied between $r_s=.352$ (Sentimental) and $r_s=.496$ (Tension). Difference between the performances obtained with the full test set to that with only one term for each track was on average $\Delta r_s=.132$, $.086\leq \Delta r_s\leq.151$, which is not a drastic drop considering that the prediction based on one term alone deals with a lot less track-level information.

These results suggest that ACT prediction is robust against the low number of track-level mood tags. Based on the results, we estimate that the correlations of $r_s=.433$ between the perceived mood and ACT mood predictions extend to the large set of almost \textit{500,000} tracks extracted from Last.fm. This gives positive implications for utilizing sparse but abundant social tags to manage and retrieve music.

\section{Conclusions}\label{sec:conclusions}

This paper marks the first systematic assessment of the potential of social tags at capturing mood information in music. We used large-scale analysis of social tags coupled with existing emotion models to construct robust music mood prediction.

We proposed a novel technique called Affective Circumplex Transformation to represent mood terms and tracks in a space of Valence, Arousal and Tension. Use of the dimensional emotion model to represent moods was supported by our analysis of the structure of the tag data. ACT outperformed the baseline techniques at predicting listener ratings of moods in a separate test set of tracks spanning multiple genres. Furthermore, the results showed that mood prediction with ACT is robust against the low number of track-level mood tags, and suggested that moderate to good fit with the dimensional emotion model can be achieved in extremely large data sets.

The present study facilitates information retrieval according to mood, assists in building large-scale mood data sets for music research, gives new ways to assess the validity of emotion models on large data relevant to current music listening habits, and makes large data available for training models that automatically annotate music based on audio.

A limitation of the present study is the possible discrepancy between track-level and clip-level moods, which may have reduced the prediction rates presented. This is because tags associated to full tracks may not adequately describe the representative clip rated in the listening experiment. Moreover, further research is needed to assess the mood-related associations of music genres and genre-related implications to mood modeling.

The implications of the present study also extend to mood mining in other online content, as it was shown that domain knowledge of moods is highly beneficial to semantic computing. Moreover, the techniques developed here can be applied to social tags as well as to other types of textual data.

\ifCLASSOPTIONcompsoc
  \section*{Acknowledgments}
\else
  \section*{Acknowledgment}
\fi

The authors wish to thank the reviewers for their invaluable feedback and suggestions.

\ifCLASSOPTIONcaptionsoff
  \newpage
\fi



\bibliographystyle{IEEEtran}

\begin{thebibliography}{10}
\providecommand{\url}[1]{#1}
\csname url@samestyle\endcsname
\providecommand{\newblock}{\relax}
\providecommand{\bibinfo}[2]{#2}
\providecommand{\BIBentrySTDinterwordspacing}{\spaceskip=0pt\relax}
\providecommand{\BIBentryALTinterwordstretchfactor}{4}
\providecommand{\BIBentryALTinterwordspacing}{\spaceskip=\fontdimen2\font plus
\BIBentryALTinterwordstretchfactor\fontdimen3\font minus
  \fontdimen4\font\relax}
\providecommand{\BIBforeignlanguage}[2]{{%
\expandafter\ifx\csname l@#1\endcsname\relax
\typeout{** WARNING: IEEEtran.bst: No hyphenation pattern has been}%
\typeout{** loaded for the language `#1'. Using the pattern for}%
\typeout{** the default language instead.}%
\else
\language=\csname l@#1\endcsname
\fi
#2}}
\providecommand{\BIBdecl}{\relax}
\BIBdecl

\bibitem{Abbasi2008}
A.~Abbasi, H.~Chen, S.~Thoms, and T.~Fu, ``Affect analysis of web forums and
  blogs using correlation ensembles,'' \emph{IEEE Transactions on Knowledge and
  Data Engineering}, vol.~20, no.~9, pp. 1168 --1180, sept. 2008.

\bibitem{Nguyen2013}
T.~Nguyen, D.~Phung, B.~Adams, and S.~Venkatesh, ``Mood sensing from social
  media texts and its applications,'' \emph{Knowledge and Information Systems},
  pp. 1--36, 2013.

\bibitem{Schmidt2009}
S.~Schmidt and W.~G. Stock, ``Collective indexing of emotions in images. a
  study in emotional information retrieval,'' \emph{Journal of the American
  Society for Information Science and Technology}, vol.~60, no.~5, pp.
  863--876, 2009.

\bibitem{Bao2012}
S.~Bao, S.~Xu, L.~Zhang, R.~Yan, Z.~Su, D.~Han, and Y.~Yu, ``Mining social
  emotions from affective text,'' \emph{IEEE Transactions on Knowledge and Data
  Engineering}, vol.~24, no.~9, pp. 1658 --1670, sept. 2012.

\bibitem{Juslin2009handbook}
P.~N. Juslin and J.~A. Sloboda, \emph{Handbook of music and emotion: Theory,
  research, applications}.\hskip 1em plus 0.5em minus 0.4em\relax Oxford
  University Press, 2009.

\bibitem{Golder2006}
S.~A. Golder and B.~A. Huberman, ``Usage patterns of collaborative tagging
  systems,'' \emph{Journal of Information Science}, vol.~32, no.~2, pp.
  198--208, April 2006.

\bibitem{Deerwester1990}
S.~Deerwester, S.~T. Dumais, G.~W. Furnas, and T.~K. Landauer, ``Indexing by
  latent semantic analysis,'' \emph{Journal of the American Society for
  Information Science}, vol.~41, no.~6, pp. 391--407, 1990.

\bibitem{Heymann2008}
P.~Heymann, D.~Ramage, and H.~Garcia-Molina, ``Social tag prediction,'' in
  \emph{Proceedings of the 31st annual international ACM SIGIR conference on
  Research and development in information retrieval}.\hskip 1em plus 0.5em
  minus 0.4em\relax ACM, 2008, pp. 531--538.

\bibitem{Song2011}
Y.~Song, L.~Zhang, and C.~L. Giles, ``Automatic tag recommendation algorithms
  for social recommender systems,'' \emph{ACM Transactions on the Web (TWEB)},
  vol.~5, no.~1, p.~4, 2011.

\bibitem{Garcia-Silva2012}
A.~Garcia-Silva, O.~Corcho, H.~Alani, and A.~Gomez-Perez, ``Review of the state
  of the art: Discovering and associating semantics to tags in folksonomies,''
  \emph{The Knowledge Engineering Review}, vol.~27, no.~01, pp. 57--85, 2012.

\bibitem{Bischoff2008}
K.~Bischoff, C.~S. Firan, W.~Nejdl, and R.~Paiu, ``Can all tags be used for
  search?'' in \emph{Proceedings of the 17th ACM conference on Information and
  knowledge management}.\hskip 1em plus 0.5em minus 0.4em\relax ACM, 2008, pp.
  193--202.

\bibitem{lamere2008}
P.~Lamere, ``Social tagging and music information retrieval,'' \emph{Journal of
  New Music Research}, vol.~37, no.~2, pp. 101--114, 2008.

\bibitem{Levy2007}
M.~Levy and M.~Sandler, ``A semantic space for music derived from social
  tags,'' in \emph{Proceedings of 8th International Conference on Music
  Information Retrieval (ISMIR)}, 2007.

\bibitem{laurier2009b}
C.~Laurier, M.~Sordo, J.~Serra, and P.~Herrera, ``Music mood representations
  from social tags,'' in \emph{Proceedings of 10th International Conference on
  Music Information Retrieval (ISMIR)}, 2009, pp. 381--86.

\bibitem{Russell1980}
J.~A. Russell, ``A circumplex model of affect,'' \emph{Journal of Personality
  and Social Psychology}, vol.~39, no.~6, pp. 1161--1178, 1980.

\bibitem{Zentner2008}
M.~Zentner, D.~Grandjean, and K.~Scherer, ``Emotions evoked by the sound of
  music: Characterization, classification, and measurement,'' \emph{Emotion},
  vol.~8, no.~4, pp. 494--521, 2008.

\bibitem{Ekman1992}
P.~Ekman, ``An argument for basic emotions,'' \emph{Cognition \& Emotion},
  vol.~6, pp. 169--200, 1992.

\bibitem{Eerola2012}
T.~Eerola and J.~K. Vuoskoski, ``A review of music and emotion studies:
  Approaches, emotion models and stimuli,'' \emph{Music Perception}, vol.~30,
  no.~3, pp. 307--340, 2012.

\bibitem{Salton1975}
G.~Salton, A.~Wong, and C.~S. Yang, ``A vector space model for automatic
  indexing,'' \emph{Communications of the ACM}, vol.~18, no.~11, pp. 613--620,
  Nov. 1975.

\bibitem{Lee2001}
D.~Seung and L.~Lee, ``Algorithms for non-negative matrix factorization,''
  \emph{Advances in neural information processing systems}, vol.~13, pp.
  556--562, 2001.

\bibitem{Hofmann2001}
T.~Hofmann, ``Unsupervised learning by probabilistic latent semantic
  analysis,'' \emph{Machine learning}, vol.~42, no. 1-2, pp. 177--196, 2001.

\bibitem{Law2010}
E.~Law, B.~Settles, and T.~Mitchell, ``Learning to tag from open vocabulary
  labels,'' in \emph{Machine Learning and Knowledge Discovery in Databases},
  ser. Lecture Notes in Computer Science, J.~Balc{\'a}zar, F.~Bonchi,
  A.~Gionis, and M.~Sebag, Eds.\hskip 1em plus 0.5em minus 0.4em\relax Springer
  Berlin Heidelberg, 2010, vol. 6322, pp. 211--226.

\bibitem{Zhou2011}
N.~Zhou, W.~Cheung, G.~Qiu, and X.~Xue, ``A hybrid probabilistic model for
  unified collaborative and content-based image tagging,'' \emph{IEEE
  Transactions on Pattern Analysis and Machine Intelligence}, vol.~33, no.~7,
  pp. 1281--1294, 2011.

\bibitem{Peng2011}
J.~Peng, D.~D. Zeng, and Z.~Huang, ``Latent subject-centered modeling of
  collaborative tagging: An application in social search,'' \emph{ACM
  Transactions on Management Information Systems}, vol.~2, no.~3, pp.
  15:1--15:23, Oct. 2008.

\bibitem{Wetzker2009}
R.~Wetzker, W.~Umbrath, and A.~Said, ``A hybrid approach to item recommendation
  in folksonomies,'' in \emph{Proceedings of the WSDM'09 Workshop on Exploiting
  Semantic Annotations in Information Retrieval}.\hskip 1em plus 0.5em minus
  0.4em\relax ACM, 2009, pp. 25--29.

\bibitem{Levy2008}
M.~Levy and M.~Sandler, ``Learning latent semantic models for music from social
  tags,'' \emph{Journal of New Music Research}, vol.~37, no.~2, pp. 137--150,
  2008.

\bibitem{Rakesh2009}
R.~Peter, G.~Shivapratap, G.~Divya, and K.~Soman, ``Evaluation of svd and nmf
  methods for latent semantic analysis,'' \emph{International Journal of Recent
  Trends in Engineering}, vol.~1, no.~3, pp. 308--310, 2009.

\bibitem{Calvo2012}
R.~A. Calvo and S.~Mac~Kim, ``Emotions in text: Dimensional and categorical
  models,'' \emph{Computational Intelligence}, 2012.

\bibitem{Scherer1984}
K.~R. Scherer, \emph{Emotion as a multicomponent process: A model and some
  cross-cultural data}.\hskip 1em plus 0.5em minus 0.4em\relax Beverly Hills:
  CA: Sage, 1984, pp. 37--63.

\bibitem{Thayer1989}
R.~E. Thayer, \emph{The Biopsychology of Mood and Arousal.}\hskip 1em plus
  0.5em minus 0.4em\relax Oxford University Press, New York, USA, 1989.

\bibitem{Juslin2004}
P.~Juslin and P.~Laukka, ``Expression, perception, and induction of musical
  emotions: A review and a questionnaire study of everyday listening,''
  \emph{Journal of New Music Research}, vol.~33, pp. 217--238, 2004.

\bibitem{juslin2011emotional}
P.~Juslin, S.~Liljestr{\"o}m, P.~Laukka, D.~V{\"a}stfj{\"a}ll, and
  L.~Lundqvist, ``Emotional reactions to music in a nationally representative
  sample of swedish adults prevalence and causal influences,'' \emph{Musicae
  scientiae}, vol.~15, no.~2, pp. 174--207, 2011.

\bibitem{zentner2010a}
M.~R. Zentner and T.~Eerola, \emph{Handbook of Music and Emotion}.\hskip 1em
  plus 0.5em minus 0.4em\relax Boston, MA: Oxford University Press, 2010, ch.
  Self-report measures and models, pp. 187--221.

\bibitem{Bradley1999}
M.~M. Bradley and P.~J. Lang, ``Affective norms for english words (anew):
  Instruction manual and affective ratings,'' Technical Report C-1, The Center
  for Research in Psychophysiology, University of Florida, Tech. Rep., 1999.

\bibitem{Saari2011}
P.~Saari, T.~Eerola, and O.~Lartillot, ``Generalizability and simplicity as
  criteria in feature selection: Application to mood classification in music,''
  \emph{IEEE Transactions on Speech and Audio Processing}, vol.~19, no.~6, pp.
  1802 --1812, aug. 2011.

\bibitem{Yang2008}
Y.~H. Yang, Y.~C. Lin, Y.~F. Su, and H.~H. Chen, ``A regression approach to
  music emotion recognition,'' \emph{IEEE Transactions on Audio, Speech, and
  Language Processing}, vol.~16, no.~2, pp. 448--457, Feb. 2008.

\bibitem{shaver1987}
P.~Shaver, J.~Schwartz, D.~Kirson, and C.~O'Connor, ``Emotion knowledge:
  further exploration of a prototype approach.'' \emph{Journal of Personality
  and Social Psychology}, vol.~52, no.~6, pp. 1061--86, 1987.

\bibitem{hevner1936}
K.~Hevner, ``Experimental studies of the elements of expression in music,''
  \emph{The American Journal of Psychology}, vol.~48, no.~2, pp. 246--268,
  1936.

\bibitem{HuDownie2007}
X.~Hu and J.~S. Downie, ``Exploring mood metadata: relationships with genre,
  artist and usage metadata,'' in \emph{Proceedings of the 8th International
  Conference on Music Information Retrieval (ISMIR)}, 2007.

\bibitem{Fellbaum1998}
C.~Fellbaum, Ed., \emph{WordNet: An Electronic Lexical Database}.\hskip 1em
  plus 0.5em minus 0.4em\relax Cambridge, MA: MIT Press, 1998.

\bibitem{Kruskal1964}
J.~Kruskal, ``\BIBforeignlanguage{English}{Multidimensional scaling by
  optimizing goodness of fit to a nonmetric hypothesis},''
  \emph{\BIBforeignlanguage{English}{Psychometrika}}, vol.~29, pp. 1--27, 1964.

\bibitem{Becavin2011}
C.~B{\'e}cavin, N.~Tchitchek, C.~Mintsa-Eya, A.~Lesne, and A.~Benecke,
  ``Improving the efficiency of multidimensional scaling in the analysis of
  high-dimensional data using singular value decomposition,''
  \emph{Bioinformatics}, vol.~27, no.~10, pp. 1413--1421, May 2011.

\bibitem{Hopkins1954}
B.~Hopkins and J.~G. Skellam, ``A new method for determining the type of
  distribution of plant individuals,'' \emph{Annals of Botany}, vol.~18, no.~2,
  pp. 231--227, 1954.

\bibitem{lawson1990}
R.~G. Lawson and P.~C. Jurs, ``New index for clustering tendency and its
  application to chemical problems,'' \emph{Journal of Chemical Information and
  Computer Sciences}, vol.~30, no.~1, pp. 36--41, 1990.

\bibitem{Gower2004}
J.~C. Gower and G.~B. Dijksterhuis, \emph{Procrustes problems}.\hskip 1em plus
  0.5em minus 0.4em\relax Oxford University Press Oxford, 2004, vol.~3.

\bibitem{Mandel2011}
M.~I. Mandel, R.~Pascanu, D.~Eck, Y.~Bengio, L.~M. Aiello, R.~Schifanella, and
  F.~Menczer, ``Contextual tag inference,'' \emph{{ACM} Transactions on
  Multimedia Computing, Communications and Applications}, vol.~7S, no.~1, pp.
  32:1--32:18, October 2011.

\bibitem{Hu:08}
X.~Hu, J.~S. Downie, C.~Laurier, M.~Bay, and A.~F. Ehmann, ``The 2007 mirex
  audio mood classification task: Lessons learned,'' in \emph{Proceedings of
  9th International Conference on Music Information Retrieval (ISMIR)}, 2008,
  pp. 462--467.

\bibitem{Nunnally1978}
J.~Nunnally, \emph{Psychometric theory}.\hskip 1em plus 0.5em minus 0.4em\relax
  New York: McGraw-Hill, 1978.

\bibitem{Eerola2011}
T.~Eerola and J.~Vuoskoski, ``A comparison of the discrete and dimensional
  models of emotion in music.'' \emph{Psychol. Music}, vol.~39, no.~1, pp.
  18--49, 2011.

\bibitem{eerola2011c}
T.~Eerola, ``Are the emotions expressed in music genre-specific? an audio-based
  evaluation of datasets spanning classical, film, pop and mixed genres,''
  \emph{Journal of New Music Research}, vol.~40, no.~4, pp. 349--366, 2011.

\end{thebibliography}

\begin{IEEEbiography}[{\includegraphics[width=1in,height=1.25in,clip,keepaspectratio]{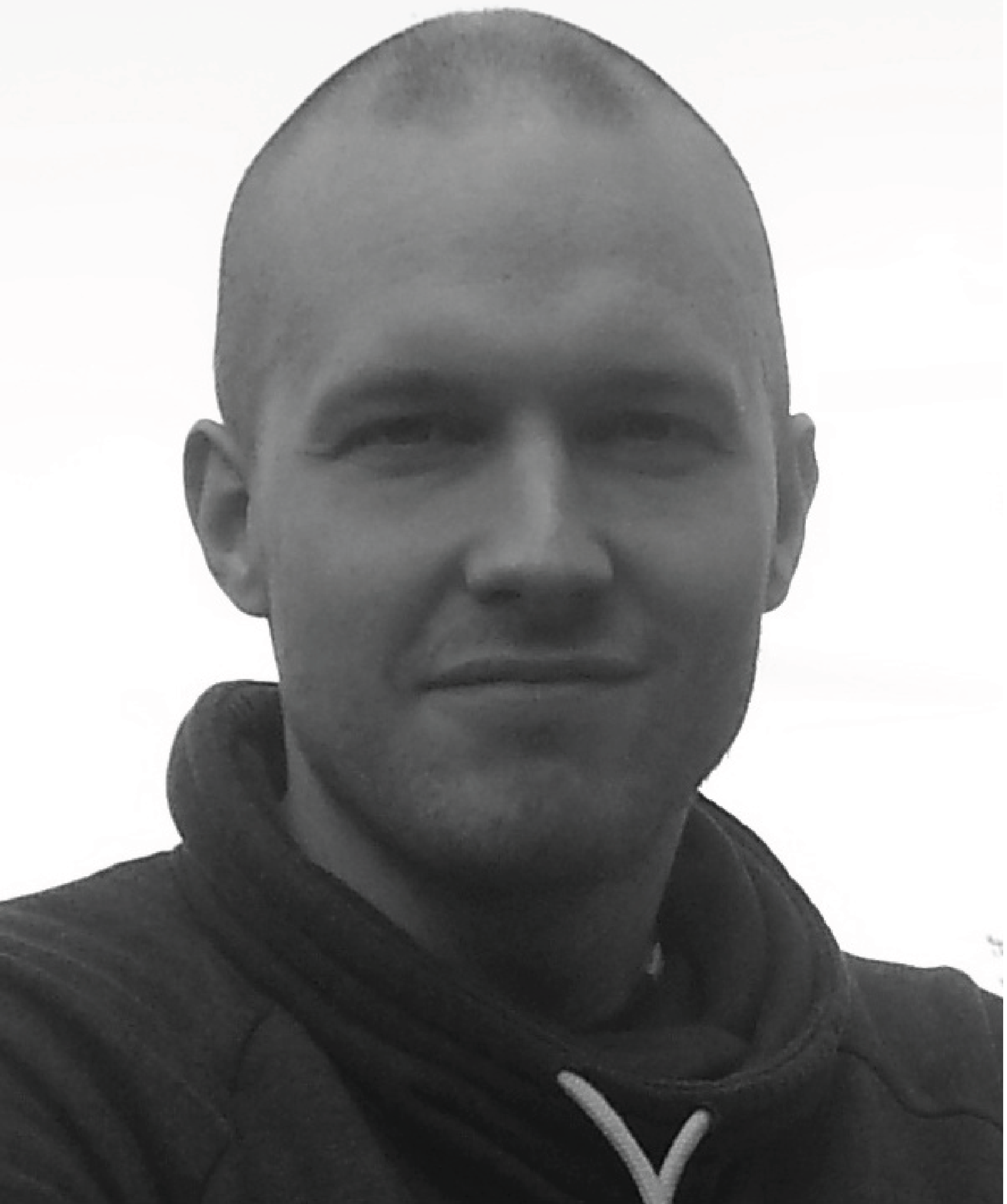}}]{Pasi Saari}
received MSc degree in Computer Science in 2008 and MA degree in Musicology in 2010 from the University of Jyv\"{a}skyl\"{a}, Finland. 

He is currently working as a Doctoral Student at the Finnish  
Centre of Excellence in Interdisciplinary Music Research within the University of Jyv\"{a}skyl\"{a}, Finland. His research interests are in semantic computing of moods in music and content-based analysis of musical audio.
\end{IEEEbiography}

\begin{IEEEbiography}[{\includegraphics[width=1in,height=1.25in,clip,keepaspectratio]{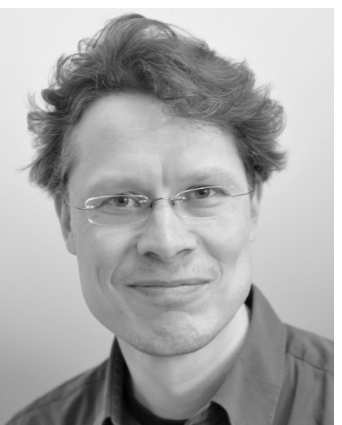}}]{Tuomas Eerola} received a PhD in musicology from the University of Jyv\"{a}skyl\"{a}, Finland and is currently a full Professor of Music at this institution. He is also affiliated with the Finnish Centre of Excellence in Interdisciplinary Music Research.

His research interest are in music cognition, particularly the perception of melody, rhythm, timbre, and induction of emotions by music.    
    
\end{IEEEbiography}




\end{document}